\documentclass[preprint,12pt]{elsarticle}

\usepackage{amsmath}
\usepackage{amsfonts}
\usepackage{amssymb}
\usepackage[utf8]{inputenc}
\usepackage[T2A]{fontenc}
\usepackage{bm}
\usepackage{graphicx}

\begin{document}

\title{
	Algebra of the spinor invariants and \\
	the relativistic hydrogen atom
%Solution of the Dirac equation \\
% with the Coulomb potential\\ within algebraic approach 
}	

\author{A.A. Eremko$^1$, L.S. Brizhik$^1$, V.M. Loktev$^{1,2}$}
\address{$^1$ Bogolyubov Institute for Theoretical Physics of the National Academy of Sciences of Ukraine \\
Metrolohichna Str., 14-b,  Kyiv, 03143, Ukraine \\
$^2$ National Technical University of Ukraine
“Igor Sikorsky Kyiv Polytechnic Institute” \\
Peremohy av., 37, Kyiv, 03056,  Ukraine}

\begin{abstract}
{ It is shown that the Dirac equation with the Coulomb potential can be solved using the algebra of the three spinor invariants of the Dirac equation without the involvement of the methods of supersymmetric quantum mechanics. The Dirac Hamiltonian is invariant with respect to the rotation transformation, which indicates the dynamical (hidden) symmetry $ SU(2) $ of the Dirac equation. The total symmetry of the Dirac equation is the symmetry   $ SO(3) \otimes SU(2) $. The generator of the $ SO(3) $ symmetry group is given by the total momentum operator, and the generator of $ SU(2) $ group is given by the rotation of the vector-states in the spinor space, determined by the Dirac, Johnson-Lippmann, and the new spinor invariants. It is shown that using algebraic approach to the Dirac problem allows one to calculate the eigenstates and eigenenergies of the relativistic hydrogen atom and reveals the fundamental role of the principal quantum number as an independent number, even though it is represented as the combination of other quantum numbers.
}

\end{abstract}

\maketitle   

\textbf{Keywords:} Dirac equation with the Coulomb potential, relativistic Kepler problem, relativistic hydrogen atom, operator invariant of the
Dirac equation, algebra of the spinor invariants, spin state, general solution of the Dirac equation.

\section{Introduction}

The problem of a particle in the central symmetric field $ V (\mathbf {r}) = V(r) \sim - 1 /r $ (Kepler problem or hydrogen atom), belongs to the class of complete integrable problems both in classical and quantum mechanics (here $ \mathbf {r} $ is the radius-vector and $r$ is its  modulus). In the quantum case this integrability is provided by the full set of independent commuting operators, which  includes operators of the Hamiltonian and integrals of motion (so called invariants), corresponding to physical variables measurable simultaneously with the energy.
 
In the non-relativistic quantum mechanics the well known solution of the Schr\"odinger equation (SE) with the Coulomb potential is provided by the invariant set $ \mathcal{H} = \lbrace \hat{H}, \hat{\mathbf{L}}^{2}, \hat{L}_{z} \rbrace $, where $ \hat{H} $ is the Hamiltonian, $ \hat{\mathbf{L}} $ is the operator of the orbital momentum and $\hat{L}_{z} $ is the operator of its projection on the polar axis $z$. The invariants $ \hat{\mathbf{L}}^{2} $ and $\hat{L}_{z} $ provide the solution in spherical coordinates; the corresponding stationary states are characterized by the set of quantum numbers $ \lbrace n,l,m \rbrace $, where $ n $ is the principal quantum number, $ l $ and $ m $ are orbital and magnetic quantum numbers, which characterize the eigenvalues of operators $ \hat{\mathbf{L}}^{2} $ and  $ \hat{L}_z $, respectively. 

It is well known that the Hamiltonian also commutes with the Laplace-Runge-Lenz vector $ \hat{\mathbf{C}} $, whose components do not commute with $ \hat{\mathbf{L}}^{2} $. The existence of several invariants with non-commuting operators is related with the so-called accidental degeneracy of the hydrogen levels with respect to number $ l $ and indicates the hidden symmetry. The exact solution of the SE is obtained by the method of variable separation: several non-commuting invariants allow one to separate variables in different coordinate systems according to various sets $ \mathcal{H} = \lbrace \hat{H}, \hat{\mathcal{I}}, \hat{L}_{z} \rbrace $. The one with $ \hat{\mathcal{I}} = \hat{\mathbf{L}}^{2} $ leads to spherical coordinate system. Separation of variables in parabolic coordinates \cite{LandL} corresponds to $ \hat{\mathcal{I}} = \hat{C}_{z} $, while in spheroidal coordinates \cite{Teller,Coulson,Mardoyan,Dullin} it is provided by the choice of the invariant $ \hat{\mathcal{I}} $ in the form of a linear combination of $ \hat{\mathbf{L}}^{2} $ and $ \hat{C}_{z} $.

In the relativistic theory propagation of a particle in the Coulomb potential is described by the Dirac equation (DE) \cite{Dirac} which also belongs to the class of complete integrable systems and admits the exact solution \cite{Dirac,Darwin,Gordon}. However, the fact that integrability of the DE is provided by the set of independent commuting operators $ \mathcal{H} = \lbrace \hat{H}_{D}, \hat{\mathbf{J}}^{2}, \hat{J}_{z},\hat{{K}} \rbrace $, is not enough elucidated. Here $ \hat{H}_{D} $ is the Dirac Hamiltonian with the Coulomb potential,  $ \mathbf{\hat{J}} = \hat{\mathbf{L}} \hat{I} + (\hbar /2) \bm{\hat{\Sigma}} $ is the total angular momentum, $ J_z $ is its projection and $ \hat{{K}} $ is the Dirac invariant. The well known Darwin solution \cite{Darwin,Gordon} of the DE corresponds namely and only to this set of invariants.
 
Unlike in the non-relativistic problems, the full set of operators in the relativistic case includes four integrals of motion because Dirac theory in a natural way takes into account the spin degree of freedom, and the corresponding stationary states in hydrogen atom are characterized by the set of quantum numbers $ \lbrace n,j,m_{j},\sigma \rbrace $, where the number $ \sigma = \pm $ describes the sign of the eigenvalue $ \kappa = \sigma (j + 1/2) $ of the invariant $ \hat{{K}} $. Besides the Dirac invariant, there exists also the Johnson-Lippmann invariant $ \hat{A} $ \cite{John-Lip}. The existence of the Johnson-Lippmann operator which does not commute with the operator $ \hat{{K}} $, explains the "accidental" degeneracy with respect to the quantum number $ \sigma $ \cite{John-Lip} and indicates the hidden symmetry. It has been shown in \cite{Sukumar,JarvSted,Dahl,Khach,Bagchi} that the energy spectrum and eigenstates of the DE with the Coulomb potential can be obtained using the supersymmetry methods of quantum mechanics. Such approach helps to understand deeper the hidden symmetry problem not only in the relativistic, but also in non-relativistic quantum mechanics when analyzing solutions of the SE. 

In our previous paper \cite{AoP22}, in addition to the Dirac and Johnson-Lippmann operators, a new invariant $\hat{I}_{BEL} $ was introduced, and the general solution of the DE for the Coulomb field has been calculated directly, using the full set of the spinor invariants that relativistically incorporate spin degree of freedom. Because operators of the spinor invariants $ \hat{K} $, $ \hat{A} $ and $ \hat{I}_{BEL} $ do not commute with each other, the  DE solution depends on the choice of the spinor invariant included into the set $ \mathcal{H} = \lbrace \hat{H}_{D}, \hat{J}_{z},\hat{\mathcal{I}}_{sp} \rbrace $. The general solution of the DE corresponds to the choice of $ \hat{\mathcal{I}}_{sp} $ as a linear combination of the three spinor invariants.

It turns out that this is not the only way to obtain the solution of the corresponding equations. The existence of additional invariants indicates not only the hidden symmetry, but also shows principal possibility to apply the group theoretical methods for constructing hydrogen atom eigenstates and eigenvalues  \cite{Pauli,Fock,Fock_Izv}. W. Pauli \cite{Pauli} noticed that the angular momentum $ \hat{\mathbf{L}} $ together with the vector $ \hat{\mathbf{C}} $ generates the hidden symmetry $ SO(4) $ and obtained hydrogen bound states spectrum by purely algebraic means. In \cite{Wipf} the celebrated results of Pauli, Fock, and others have been generalized to the hydrogen atom in arbitrary dimensions and to the corresponding supersymmetric extensions. The algebraic properties of the Coulomb potential have been discussed in \cite{Katsura} and in \cite{Khelashvili} where the energy spectrum of the DE with the Coulomb potential has been calculated, based on  Witten’s superalgebra. 

Therefore, below, by exploiting existence of the three invariants $ \hat{K} $, $ \hat{A} $ and $ \hat{I}_{BEL} $, we calculate the solution of the DE with the Coulomb potential using the algebra of the spinor invariant operators without the involvement of the supersymmetric quantum mechanics methods. As it will be shown below, this approach is not simply the alternative method to find the solution. It demonstrates the fundamental role of the principal quantum number as an intrinsic characteristic of the electron states in hydrogen atoms, not as a mere combination of the quantum numbers it is composed of, as it is usually introduced in the conventional, direct methods of solving the equations. 

\section{Invariants of the Dirac Hamiltonian and their eigenvalues}

Recall, the Dirac Hamiltonian for the external central symmetric potential, created by a point positive charge $ Ze $, has the form 
\begin{equation}
\label{H_D-V} 
\hat{H}_{D} = c\bm{\hat{\alpha}} \cdot \mathbf{\hat{p}} + m c^{2} \hat{\beta} -\frac{Ze^{2}}{r} \hat{I}_4, \quad r = \vert \mathbf{r} \vert = \sqrt{x^{2} + y^{2} + z^{2}}  ,  
\end{equation}
in which $ \hat{I}_4 $ is a unit $ 4 \times 4 $ matrix, $ \bm{\hat{\alpha}} $ and $ \hat{\beta} $ are Dirac matrices, $ m $ is particle's mass, $ \hat{\mathbf{p}} $ is momentum and $ c $ is the speed of light. 
 
The total momentum 
\begin{equation}
\label{J_tot}
\mathbf{\hat{J}} = \mathbf{L} + \frac{1}{2} \hbar \bm{\hat{\Sigma}} 
\end{equation} 
is the invariant of this Hamiltonian. The invariants of the DE are given by the operators:\\   
1) one of the total momentum components, whose direction is usually chosen along the polar (or quantization) axis  $z$,
\begin{equation}
\label{J_j}
\hat{J}_{z} = \hat{L}_{z} + \frac{1}{2} \hbar \hat{\Sigma}_{z} ;
\end{equation} 
2) absolute value of the total momentum  $ \mathbf{\hat{J}} $,  
\begin{equation}
\label{J^2}
\mathbf{\hat{J}}^{2} = \hat{J}_{x}^{2} + \hat{J}_{y}^{2} + \hat{J}_{z}^{2} = \mathbf{\hat{L}}^{2} \hat{I}_4 + \hbar \bm{\hat{\Sigma}}\cdot \mathbf{\hat{L}} + \frac{3}{4} \hbar^{2} \hat{I}_4 ;
\end{equation}
3) Dirac invariant
\begin{equation}
\label{Ksc} 
\hat{K} = \bm{\hat{\Omega}} \cdot \hat{\mathbf{L}} + \hbar \hat{\beta};
\end{equation}
4) Johnson-Lippmann invariant   
\begin{equation}
\label{invJ-L} 
\hat{A} = \frac{mZe^{2}}{r} \bm{\hat{\Sigma}} \cdot \mathbf{r} - \bm{\hat{\Omega}} \cdot \frac{1}{2} \left( \mathbf{\hat{p}} \times \hat{\mathbf{L}} - \hat{\mathbf{L}} \times \mathbf{\hat{p}} \right) - \frac{Ze^{2}}{cr} \left( \bm{\hat{\Gamma}} \cdot \hat{\mathbf{L}} + \hbar \hat{\rho}_{2} \right) ;
\end{equation} 
5) and new invariant, introduced in Ref. \cite{AoP22}, 
\begin{equation}
\label{invBEL} 
\hat{I}_{BEL} = \frac{1}{2i}\left[ \hat{K}, \hat{A} \right] .
\end{equation}

Here we use the conventional representation of the Dirac matrices, as we used in   \cite{AoP15}, namely 
\[
\bm{\hat{\alpha}} = \left( 
\begin{array}{cc}
0 & \bm{\hat{\sigma}} \\ \bm{\hat{\sigma}} & 0
\end{array} \right) , \quad 
\hat{\beta} = \left( \begin{array}{cc}
\hat{I} & 0 \\ 0 & -\hat{I}
\end{array}  \right) 
, \quad 
\bm{\hat{\Sigma}} = \left( 
\begin{array}{cc}
\bm{\hat{\sigma}} & 0 \\ 0 & \bm{\hat{\sigma}}
\end{array} \right)
\]
and notations of the Hermitian matrices which are obtained as the products of the four Dirac matrices  $ \hat{\alpha}_{j} $ ($ j=x,y,z $) and $ \hat{\beta} $:
\[
\begin{array}{c}
\hat{\rho}_{1} = -i \hat{\alpha}_{x} \hat{\alpha}_{y} \hat{\alpha}_{z} = \left( \begin{array}{cc}
0 & \hat{I}_{2} \\ \hat{I}_{2} & 0
\end{array}  \right), \quad \hat{\rho}_{2} = -i\hat{\beta}\hat{\rho}_{1} = \left( \begin{array}{cc}
0 & -i \hat{I}_{2} \\ i \hat{I}_{2} & 0
\end{array}  \right) , \\
\bm{\hat{\Gamma}} = -i \hat{\beta} \bm{\hat{\alpha}} = \left( 
\begin{array}{cc}
0 & -i\bm{\hat{\sigma}} \\ i\bm{\hat{\sigma}} & 0
\end{array} \right) , \quad \bm{\hat{\Omega}} = \hat{\beta} \bm{\hat{\Sigma}} = \left( 
\begin{array}{cc}
\bm{\hat{\sigma}} & 0 \\ 0 & -\bm{\hat{\sigma}}
\end{array} \right) ,
\end{array} 
\]
where $ \hat{I}_{2} $ is a unit $2 \times 2 $-matrix, and $ \bm{\hat{\sigma}} $ is Pauli vector-matrix. Matrices $ \bm{\hat{\alpha}} $, $ \bm{\hat{\Gamma}} $, $ \bm{\hat{\Omega}} $, $ \bm{\hat{\Sigma}} $, $ \hat{\rho}_{1} $, $ \hat{\rho}_{2} $, $ \hat{\rho}_{3} $ and the unit $4 \times 4 $ matrix compose the group of 16 linearly independent Hermitian matrices. Any $4 \times 4 $ matrix can be unambiguously represented as a linear combination of these matrices. Recall, Dirac matrices in the form $ \hat{\rho}_{1} $, $ \hat{\rho}_{2} $ and $ \hat{\rho}_{3} = \hat{\beta} $ are used sometimes in the literature, in particular, in Ref. \cite{John-Lip,Dahl}.

It is worth to note that the invariant (\ref{invBEL}) is, in fact, the consequence of the classical Poisson theorem, according to which the Poisson bracket of the two integrals of motion (in quantum mechanics it is their commutator) is also the integral of motion. The Poisson theorem far not always provides qualitatively new consequences and gives a new independent integral of motion. It can give a trivial operator (such as a constant) or a function of the known integrals of motion. As an example we write down the operator 
\[
\hat{Y} = \bm{\hat{\Sigma}}\cdot \mathbf{\hat{L}} + \frac{1}{mc} \bm{\hat{\Gamma}} \cdot \frac{1}{2} \left( \mathbf{\hat{p}} \times \hat{\mathbf{L}} - \hat{\mathbf{L}} \times \mathbf{\hat{p}} \right) - \frac{Ze^{2}}{mc^{2} r} \left( \bm{\hat{\Omega}} \cdot \mathbf{\hat{L}} + \hbar \hat{\beta} \right),
\]
which commute with the Hamiltonian (\ref{H_D-V}), but are not independent and do not contain any new physical information since it commute with the invariant (\ref{Ksc}) and they have common system of eigen bispinors. However, it is easy to check that operator (\ref{invBEL}) is an independent operator non commuting with the invariant (\ref{Ksc}) and, thus, ought to be taken into account in  solving the DE.  

\section{Stationary states of the Dirac equation }

Solutions of the DE $ \hat{H}_{D} \vert \Psi \rangle = E \vert \Psi \rangle $ with the Hamiltonian (\ref{H_D-V}), which describe its eigenstates, are characterized by the set of quantum numbers, that follow from the joint solution of the system 
\begin{equation}
\label{state} 
\begin{array}{c}
\hat{H}_{D}\vert \Psi \rangle = E \vert \Psi \rangle , \\ 
\mathbf{\hat{J}}^{2} \vert \Psi \rangle = \hbar^{2} j\left(j + 1 \right) \vert \Psi \rangle , \\
\hat{J}_{z} \vert \Psi \rangle = \hbar m_{j} \vert \Psi \rangle , \\
\hat{{\cal I}}_{inv} \vert \Psi \rangle = \epsilon_{inv} \vert \Psi \rangle ,
\end{array}
\end{equation}
where the operators  $ \hat{J}_{z} $ and $ \mathbf{\hat{J}}^{2} $ are defined in Eqs. (\ref{J_j})-(\ref{J^2}), and $ \hat{{\cal I}}_{inv} $ is one of the operator invariants (\ref{Ksc})-(\ref{invBEL}), $ \hat{{\cal I}}_{inv}= \hat{K},\,\hat{A},\, \hat{I}_{BEL} $. It is clear that the set of the quantum numbers which are defined by the system   (\ref{state}), that includes the full set of spinor invariants, is complete. Here invariants   (\ref{Ksc}), (\ref{invJ-L}) and (\ref{invBEL}) commute with the Hamiltonian $ \hat{H}_{D} $, as well as with the operator  $ \mathbf{\hat{J}}^{2} $ and the total momentum components, but they do not commute between themselves. Therefore, they can not have common system of eigenfunctions, and, being defined by Eqs.  (\ref{state}),  each system directly depends on the choice of the particular invariant from the set $ \hat{{\cal I}}_{inv}$.

 Thus, even before finding solutions of the DE, one can state that stationary states of a particle in the Coulomb field have certain values of the  energy
 $ E $, total momentum, its projection on the polar axis, and depend on the given spinor invariant  $ \hat{{\cal I}}_{inv} $. In other words, eigenvalues of the invariants supplement the quantum numbers which characterize stationary states of the system,  $ \vert \Psi \rangle $. Some of these numbers are commonly known, in particular, the eigenvalues of the operators  $ \mathbf{\hat{J}}^{2} $ and $ \hat{J}_{z} $ in the system (\ref{state}) can be easily calculated algebraically, taking into account commutation relations between the components of the operator  $ \mathbf{\hat{J}} $. The set of the quantum numbers includes positive semi-integer numbers $ j = 1/2,3/2,\ldots $ and semi-integer numbers $ m_{j} $ ($ -j \leq m_{j} \leq j $) which define the values of the total momentum and its projection on the polar axis and are  independent on the choice of the particular operator $ \hat{{\cal I}}_{inv} $. Therefore, below we shall define vector-state (bispinor) in Eqs. (\ref{state}) as $ \vert \Psi \rangle = \vert E ,j,m_{j},\epsilon_{inv} \rangle $.

 Invariants  eigenvalues   can be also found without solving the corresponding equations, using  squares of the invariants  \cite{Dahl}. It is easy to calculate the squares of the corresponding three operators:
\begin{equation}
\label{K^2} 
\hat{K}^{2} = \mathbf{\hat{J}}^{2} + \frac{\hbar^{2}}{4} \hat{I_4} , \quad \hat{A}^{2} = m^{2}Z^{2}e^{4} \hat{I_4} + \frac{\hat{K}^{2}}{c^{2}} \left( \hat{H}_{D}^{2} - m^{2} c^{4} \hat{I_4} \right) , \quad \hat{I}_{BEL}^{2} = \hat{K}^{2} \hat{A}^{2} .
\end{equation}

One can see that the eigenstates of Eqs. (\ref{state}) are also eigenvectors of the operators   $ \hat{K}^{2} $, $ \hat{A}^{2} $ and $ \hat{I}_{BEL}^{2} $ independent of the concrete invariant  in the last equation in system (\ref{state}). Actions of the operators  (\ref{K^2}) on the eigenstates determine  eigenvalue squares $ \epsilon_{K}^{2} $, $ \epsilon_{A}^{2} $, $ \epsilon_{I_{BEL}}^{2} $ of the operators $ \hat{K} $, $ \hat{A} $ and $ \hat{I}_{BEL}$, respectively. Taking into account the first three equalities in Eq. (\ref{state}), one gets 
\[
\epsilon_{K}^{2} = \hbar^{2} \left( j + 1/2 \right)^{2} = \hbar^{2} \kappa_{j}^{2} , \: \epsilon_{A}^{2} = m^{2}Z^{2}e^{4} + \frac{\hbar^{2}\kappa_{j}^{2}}{c^{2}}  \left( E^{2} - m^{2} c^{4}\right) = \mathrm{a}_{E,j}^{2} , \:  \epsilon_{I_{BEL}}^{2} = \hbar^{2} \kappa_{j}^{2} \mathrm{a}_{E,j}^{2}. 
\]

Introducing the value  
\begin{equation}
\label{kappa_j} 
\kappa_{j} = j + \frac{1}{2} , 
\end{equation}
which takes natural numbers starting from 1, since  $ j $ is positive,  and  the notation
\begin{equation}
\label{a_E,j_1} 
\mathrm{a}_{E,j} = \sqrt{m^{2}Z^{2}e^{4} + \frac{\hbar^{2}\kappa_{j}^{2}}{c^{2}}  \left( E^{2} - m^{2} c^{4}\right) } \, ,
\end{equation}
we find that the values 
\begin{equation}
\label{inv-eigvals_1} 
\epsilon_{K} = \pm \hbar \kappa_{j} , \quad \epsilon_{A} = \pm \mathrm{a}_{E,j} , \quad \epsilon_{I_{BEL}} = \pm \hbar \kappa_{j} \mathrm{a}_{E,j} 
\end{equation}
are the eigenvalues of the corresponding invariants (\ref{Ksc})-(\ref{invBEL}). 

So, at the given energy and total momentum,  each invariant takes two values of opposite signs. This means that together with the quantum numbers  $ j $ and $ m_{j} $, eigen vectors should be also  defined by the sign of the corresponding spinor invariant,  $ \sigma = \pm $. Hence, the number  $ \sigma $  must be included in the full set of the quantum numbers that characterize the stationary states,  $ \vert \Psi \rangle = \vert E ,j,m_{j},\sigma \rangle_{inv} $. As the result, the given invariant determines the explicit expression and the form of the eigen vectors of the DE, and this is controlled by the index  $ \lbrace inv \rbrace $. 

\section{Algebra of the invariants}

Let us consider algebraic properties of the given above integrals of motion of the DE (\ref{J_j})-(\ref{invBEL}) which we will use below. It is easy to see that each invariant including the Hamiltonian (\ref{H_D-V}), has its own dimensionality.  Nevertheless, it is often  convenient to work with the dimensionless values, so we introduce the operators 
\begin{equation}
\label{dimless} 
\begin{array}{c}
\hat{\mathcal{H}} = \hat{H}_{D}/mc^{2}, \quad \hat{\mathcal{J}}_{z} = \hat{J}_{z}/\hbar , \quad \hat{\mathcal{J}}^{2} = \mathbf{\hat{J}}^{2} /\hbar^{2} , \\
\hat{\mathcal{K}} = \hat{K}/\hbar , \quad \hat{\mathcal{A}} = \hat{A} /\left( mZe^{2} \right) , \quad \hat{\mathcal{I}} = \hat{I}_{BEL}/\left( \hbar mZe^{2} \right) ,
\end{array}
\end{equation}
which have dimensionless eigenvalues 
 $ \varepsilon = E/mc^{2} $, $ m_{j} $, $ j\left(j + 1 \right) $ and, according to  (\ref{inv-eigvals_1}), $ \epsilon_{\mathcal{K}} = \pm \kappa_{j} $, $ \epsilon_{\mathcal{A}} = \pm a_{\varepsilon,j} $, $ \epsilon_{\mathcal{I}} = \pm \kappa_{j} a_{\varepsilon,j} $, where  $ \kappa_{j} $ is defined in Eq. (\ref{kappa_j}). Here 
\begin{equation}
\label{a_E,j} 
a_{\varepsilon,j} = \sqrt{1 + \frac{\kappa_{j}^{2}}{Z^{2}\alpha^{2}}\left(\varepsilon^{2} - 1 \right) }    
\end{equation}
is the dimensionless eigenvalue  (\ref{a_E,j_1}), and parameter  $ \alpha = e^{2}/\hbar  c $ is \textit{Sommerfeld's fine structure constant}. 
 
With these notations and index $ \sigma = \pm $ included in the full set of quantum numbers, Eqs. (\ref{state}) can be rewritten as
\begin{equation}
\label{state_3} 
\begin{array}{c}
\hat{\mathcal{H}}\vert \varepsilon ,j, m_{j},\sigma \rangle_{inv} = \varepsilon \vert \varepsilon ,j, m_{j},\sigma \rangle_{inv} , \\ 
\hat{\mathcal{J}}^{2} \vert \varepsilon ,j, m_{j},\sigma \rangle_{inv} = j\left(j + 1 \right) \vert \varepsilon ,j, m_{j},\sigma \rangle_{inv} , \\
\hat{\mathcal{J}}_{z} \vert \varepsilon ,j, m_{j},\sigma \rangle_{inv} = m_{j} \vert \varepsilon ,j, m_{j},\sigma \rangle_{inv} , \\
\hat{{\cal I}}_{inv} \vert \varepsilon ,j, m_{j},\sigma \rangle_{inv} = \sigma \epsilon_{inv} \vert \varepsilon ,j, m_{j},\sigma \rangle_{inv} ,
\end{array}
\end{equation} 
where $ \epsilon_{inv} $ is the absolute value of the eigenvalue of the operator $ \hat{{\cal I}}_{inv} = \hat{\mathcal{K}} \, , \hat{\mathcal{A}} $ or $ \hat{\mathcal{I}} $, i.e., $ \epsilon_{inv}  = \kappa_{j} ,\,  a_{\varepsilon,j} $ or $ \kappa_{j} a_{\varepsilon,j} $.

The squares of the introduced operators  $ \hat{\mathcal{K}} $, $ \hat{\mathcal{A}} $ and $ \hat{\mathcal{I}} $, similar to the squares of the corresponding dimensional ones, are functions of the operators of the first two invariants in Eq. (\ref{state_3}):
\begin{equation} 
\label{G_1,2,3} 
\hat{\mathcal{K}}^{2} = \hat{I}_4/4 + \hat{\mathcal{J}}^{2} \equiv \hat{G}_{1}, \; \hat{\mathcal{A}}^{2} = \hat{I}_4 + \frac{1}{Z^{2} \alpha^{2}} \hat{G}_{1} \left( \hat{\mathcal{H}}^{2} - 1 \right) \equiv \hat{G}_{2} , \; \hat{\mathcal{I}}^{2} = \hat{G}_{1} \hat{G}_{2} \equiv \hat{G}_{3} .
\end{equation} 
Therefore, independent of the invariant in the last equation of system (\ref{state_3}), they have certain values in the space of the vector-states 
 $ \vert \varepsilon ,j, m_{j},\sigma \rangle_{inv} $: 
\begin{equation}
\label{eigenG_123} 
\begin{array}{c}
\hat{G}_{1} \vert \varepsilon ,j, m_{j},\sigma \rangle_{inv} = \kappa_{j}^{2} \vert \varepsilon ,j, m_{j},\sigma \rangle_{inv} , \quad \hat{G}_{2} \vert \varepsilon ,j, m_{j},\sigma \rangle_{inv} = a_{\varepsilon,j}^{2} \vert \varepsilon ,j, m_{j},\sigma \rangle_{inv} , \\
\hat{G}_{3} \vert \varepsilon ,j, m_{j},\sigma \rangle_{inv} = \kappa_{j}^{2} a_{\varepsilon,j}^{2} \vert \varepsilon ,j, m_{j},\sigma \rangle_{inv} ,
\end{array}  
\end{equation}
in which the values $ \kappa_{j} $ and $ a_{\varepsilon,j} $ are defined in Eqs.(\ref{kappa_j}) and  (\ref{a_E,j}), respectively.

Operators $ \hat{K} $ and $ \hat{A} $ anticommute with each other. It is easy to show that  operator  $ \hat{I}_{BEL} $ (\ref{invBEL}) also anticommutes with them. The anticommutating properties are preserved for operators  $ \hat{\mathcal{K}} $, $ \hat{\mathcal{A}} $ and $ \hat{\mathcal{I}} $: 
\[
\left\lbrace \hat{\mathcal{K}},\hat{\mathcal{A}} \right\rbrace = \left\lbrace \hat{\mathcal{I}},\hat{\mathcal{K}} \right\rbrace = \left\lbrace \hat{\mathcal{A}},\hat{\mathcal{I}} \right\rbrace  = 0 .
\]
According to the definition (\ref{invBEL}) and notations (\ref{G_1,2,3}), the following commutation relations are valid:
\begin{equation}
\label{commut} 
\left[ \hat{\mathcal{K}},\hat{\mathcal{A}} \right] = 2i \hat{\mathcal{I}}, \quad 
\left[ \hat{\mathcal{I}}, \hat{\mathcal{K}} \right] = 2i \hat{G}_{1} \hat{\mathcal{A}} , \quad 
\left[ \hat{\mathcal{A}}, \hat{\mathcal{I}} \right] = 2i \hat{G}_{2} \hat{\mathcal{K}}  .
\end{equation}

These commutators  do not lead to new independent operators, they include the operators $ \hat{{\cal I}}_{inv} $ and the operator functions of the integrals of motion $ \hat{\mathcal{H}} $ and $ \hat{\mathcal{J}}^{2} $, with which these operators commute. Therefore, the three operators  $ \hat{\mathcal{K}} $, $ \hat{\mathcal{A}} $ and $ \hat{\mathcal{I}} $ form the group whose algebra  is defined by the commutators (\ref{commut}) in which operator $ \hat{G}_{\alpha} $, $(\alpha =1,2,3)$ commutes with each operator of the group and plays the role of the 'structural constant'. Operators $ \hat{G}_{\alpha} $ in the space of eigenstates  (\ref{state_3}) reduce to their eigenvalues  (\ref{eigenG_123}).

Let us introduce the operator
\begin{equation}
\label{calC} 
\hat{\mathcal{C}} = \hat{\mathcal{K}} + \hat{\mathcal{A}} + \hat{\mathcal{I}} ,
\end{equation}
the square of which has the form 
\begin{equation}
\label{C^2} 
\hat{\mathcal{C}}^{2} = \hat{\mathcal{K}}^{2} + \hat{\mathcal{A}}^{2} + \hat{\mathcal{I}}^{2} .
\end{equation}
One can consider matrix  (\ref{calC}) as the vector representation in an abstract three-dimensional "Hilbert" space with the projections on the three orthogonal "directions"  $ \hat{\mathcal{K}} $, $ \hat{\mathcal{A}} $ and $ \hat{\mathcal{I}} $. These vector projections do not commute, and, hence, can not have simultaneously definite non-zero values.   

Next, operator (\ref{C^2}) commutes with the operators of each component   (\ref{calC}). Being the invariant, it has definite value simultaneously with one of its constituent operator, for instance, with $ \hat{\mathcal{K}} $: 
\[
\hat{\mathcal{K}} \vert \varepsilon ,j, m_{j},\sigma \rangle_{\mathcal{K}} = \sigma \kappa_{j} \vert \varepsilon ,j, m_{j},\sigma \rangle_{\mathcal{K}} , \quad 
\hat{\mathcal{C}}^{2} \vert \varepsilon ,j, m_{j},\sigma \rangle_{\mathcal{K}} = \lambda \vert \varepsilon ,j, m_{j},\sigma \rangle_{\mathcal{K}} . 
\]
Here, according to (\ref{G_1,2,3}) and (\ref{eigenG_123}), value $\lambda  $ is given by the relation
\begin{equation}
\label{lambda} 
\lambda = \kappa_{j}^{2} + (1 + \kappa_{j}^{2}) a_{\varepsilon,j}^{2} .
\end{equation}

Because the equality $ \hat{\mathcal{C}}^{2} - \hat{\mathcal{K}}^{2} = \hat{\mathcal{A}}^{2} + \hat{\mathcal{I}}^{2} $ is valid, one can write down the equation
\[
\left( \hat{\mathcal{C}}^{2} - \hat{\mathcal{K}}^{2} \right) \vert \varepsilon ,j, m_{j},\sigma \rangle_{\mathcal{K}} = \left( \hat{\mathcal{I}}^{2} + \hat{\mathcal{A}}^{2} \right) \vert \varepsilon ,j, m_{j},\sigma \rangle_{\mathcal{K}} .
\]
Eigenvalues $ \sigma \kappa_{j} $ of the operator $ \hat{\mathcal{K}} $ where $ \sigma = \pm $, are defined by integer numbers (\ref{kappa_j}) starting with $ \kappa_{j} = 1 $. Since the difference $ \hat{\mathcal{C}}^{2} - \hat{\mathcal{K}}^{2} $ corresponds to essentially positive operator,  possible eigenvalues of the operator $ \hat{\mathcal{K}} $ at the given eigenvalue of $ \hat{\mathcal{C}}^{2} $ are bound by the inequality $ \hat{\mathcal{K}}^{2} \leq \hat{\mathcal{C}}^{2} $ which is equivalent to the inequality $ \kappa_{j}^{2} \leq \lambda $. At the equality sign this condition determines the maximal possible value $ \kappa_{j} = \kappa_{max} $ at the given $ \lambda $. According to Eq. (\ref{lambda}) the equality $ \kappa_{max}^{2} = \lambda $ leads to the equality $ \left( 1 + \kappa_{max}^{2} \right) a_{\varepsilon,j_{max}}^{2} = 0 $, which is reduced to the condition $ a_{\varepsilon,j}^{2} = 0 $. Taking into account definition (\ref{a_E,j})), it can be written in the form
\begin{equation}
\label{eqE_n} 
 \kappa_{max}^{2} \varepsilon^{2} - \left( \kappa_{max}^{2} - Z^{2}\alpha^{2} \right) = 0  .
\end{equation} 
Since $ \kappa_{j} $ is a number of the natural series, we can define its maximal possible value at the given $ \lambda $ as $ \kappa_{max} = n $. According to definition  (\ref{kappa_j}), this equality defines maximal possible value of the orbital quantum number $ j_{max} = n - 1/2 $ at the given $ n $.

At not too large $ Z $ the inequality $ Z^{2}\alpha^{2} < 1 $ is fulfilled, and since $ \kappa_{j} \geqslant 1 $, the inequality $ \kappa_{j}^{2} - Z^{2}\alpha^{2} > 0 $, and, thus, the value  
\begin{equation}
\label{gamma_j} 
\gamma_{j} = \sqrt{\kappa_{j}^{2} - Z^{2}\alpha^{2}} 
\end{equation} 
can be introduced. According to this notation, $ \gamma_{j_{max}} = \gamma_{n} = \sqrt{n^{2} - Z^{2}\alpha^{2}} $, $ \kappa_{max}^{2} = \gamma_{n}^{2} + Z^{2}\alpha^{2} $, and Eq. (\ref{eqE_n}) gives the eigenvalues of the Hamiltonian 
\begin{equation}
\label{E_n} 
\varepsilon_{n,j_{max}}^{2} = \frac{\gamma_{n}^{2}}{\gamma_{n}^{2} + Z^{2}\alpha^{2}} , \quad \varepsilon_{n,j_{max}} = \sqrt{1 - \frac{Z^{2}\alpha^{2}}{\gamma_{n}^{2} + Z^{2}\alpha^{2}}} .
\end{equation} 

These expressions are obtained from the equality $ a_{\varepsilon,j}^{2} = 0 $ ($ \lambda = \kappa_{j}^{2} $) which corresponds to the maximal possible value $ \kappa_{j} $ that satisfies the condition $ \kappa_{j}^{2} \leq \lambda $. Then the energy (\ref{E_n}) which takes the standard form $ \varepsilon_{n} = \sqrt{1 - Z^{2}\alpha^{2}/n^{2}} $, corresponds to the eigenstate vector $ \vert \varepsilon_{n} ,j_{max}, m_{j},\sigma \rangle_{K} $. Nevertheless, at $ n > 1 $ there are other values $ j < j_{max} $ for which $ \kappa_{j} = n - n_{r} $ ($ n_{r} = 1,\ldots\, ,n-1 $) and the inequality $ \kappa_{j}^{2} < \lambda $ takes place. 

Summarizing the above, we conclude that stationary states of the DE are given by the vectors in the Hilbert space $ \vert n,j, m_{j},\sigma \rangle_{K} $ which are characterized by the set of quantum numbers $ \lbrace n,j,m_{j},\sigma \rbrace $. Here $ n = 1,2,\ldots $ attains the meaning of the \textit{principal quantum number}, the number $ j $ takes $ n $ values $ j = 1/2,\ldots\, , n - 1/2 $, $ m_{j} = \pm 1/2,\ldots\, ,\pm j $, and $ \sigma = \pm $. 

Thus, the value $ n = 1 $ corresponds to the ground state 
$ \vert 1 ,1/2, m_{j},\sigma \rangle_{K} $ with the energy $ \varepsilon_{1} $ (see Eq. (\ref{E_n})), while the values $ n > 1 $ are attributed to the excited states. For the given principal quantum number there are $ n $ states with different $ j = j_{max}-n_{r} $ ($ n_{r} = 0,1,\ldots\, ,n-1 $):
\begin{equation}
\label{eignvecs_n} 
\vert n ,j_{max}, m_{j},\sigma \rangle_{K} , \ldots \, , \vert n ,j_{max} - n_{r}, m_{j},\sigma \rangle_{K} , \ldots \, , \vert n ,1/2, m_{j},\sigma \rangle_{K} .
\end{equation}

The energy of the states with $ j = j_{max} $ (i.e., $ n_{r} = 0 $) is determined by Eq. (\ref{E_n}). For these states the equality $ a_{\varepsilon,j}^{2} = 0 $ is valid where $ a_{\varepsilon,j} $ was defined in Eq. (\ref{a_E,j}). So, we come to conclusion that the eigenvalue of the operator $ \hat{G}_{2} $ in these states is equal to zero (see (\ref{eigenG_123})) and, therefore, the eigenvalues of the invariants $ \hat{\mathcal{A}} $ and $ \hat{\mathcal{I}} $ are equal to zero as well.

Because $ n $ values of $ j $ at given $ n $ can be written as $ j = n - n_{r} - 1/2 $ with $ n_{r} = 0,1,\ldots\, ,n-1 $, the states (\ref{eignvecs_n}) can be characterized by the number  $ n_{r} $ which has the meaning of the  \textit{radial quantum number}. Respectively,  eigenvalues of the operators $ \hat{\mathcal{H}} $ and $ \hat{G}_{2} $ in the states $ \vert n ,n_{r}, m_{j},\sigma \rangle_{K} $ depend on $ n $ and $ n_{r} $, and can be written as $ \varepsilon_{n.n_{r}} $ and $ a_{n,n_{r}}^{2} $, respectively. 

While for the states with $ n_{r} = 0 $ the equalities $ a_{n,0} = 0 $ and $ \varepsilon_{n.0} = \varepsilon_{n} $ (see Eq.(\ref{E_n})) are fulfilled, $ a_{n,n_{r}} \neq 0 $ for other states. In the meantime, the general form of  $ \varepsilon_{n.n_{r}} $, which according to (\ref{E_n}) can be written down as  
\begin{equation}
\label{E-Q} 
\varepsilon_{n,n_{r}}^{2} = \frac{\gamma_{n,n_{r}}^{2}}{\gamma_{n,n_{r}}^{2} + Z^{2}\alpha^{2}} = 1 - \frac{Z^{2}\alpha^{2}}{\gamma_{n,n_{r}}^{2} + Z^{2}\alpha^{2}} ,
\end{equation}
must be such that for the state with $ n_{r} = 0 $ the equality $ a_{n,0} = 0 $ should be valid and (\ref{E-Q}) should coincide with (\ref{E_n}). 

Substitution of Eq. (\ref{E-Q}) into the expression for $ a_{n,n_{r}}^{2} $, defined by Eq.(\ref{a_E,j}), gives 
\[
a_{n,n_{r}}^{2} = \frac{\gamma_{n,n_{r}}^2 -\gamma_{j}^2 }{\gamma_{n,n_{r}}^{2} + Z^{2}\alpha^{2}} = \frac{\left( \gamma_{n,n_{r}} - \gamma_{j} \right) \left( \gamma_{n,n_{r}} + \gamma_{j} \right)}{\gamma_{n,n_{r}}^{2} + Z^{2}\alpha^{2}} .
\]
The necessary condition for the equality $ a_{n,0}^{2} = 0 $ is satisfied when $ \gamma_{n,n_{r}} - \gamma_{j} = n_{r} $, and  Eq. (\ref{E-Q}) with $ \gamma_{n,n_{r}} = n_{r} + \gamma_{j} $ gives the well known relativistic spectrum of the hydrogen-like systems  \cite{Dirac,Darwin,Gordon}.

\section{Eigen bispinors of the Dirac equation }

It is convenient to search the eigen bispinors in the space which is characterized by the  unit   matrices
\[
\hat{\mathcal{I}}_{1} = \frac{\hat{\mathcal{K}}}{\sqrt{\hat{G}_{1}}} , \quad \hat{\mathcal{I}}_{2} = \frac{\hat{\mathcal{A}}}{\sqrt{\hat{G}_{2}}} , \quad \hat{\mathcal{I}}_{3} = \frac{\hat{\mathcal{I}}}{\sqrt{\hat{G}_{3}}} , 
\]
such that 
\[\hat{\mathcal{I}}_{\alpha}^{2} = 1 , \quad  \alpha=1,2,3\]
whose algebra  
\[
\left[ \hat{\mathcal{I}}_{1},\hat{\mathcal{I}}_{2} \right] = 2i \hat{\mathcal{I}}_{3} , \quad \left[ \hat{\mathcal{I}}_{3},\hat{\mathcal{I}}_{1} \right] = 2i \hat{\mathcal{I}}_{2} , \quad \left[ \hat{\mathcal{I}}_{2},\hat{\mathcal{I}}_{3} \right] = 2i \hat{\mathcal{I}}_{1} ,
\]
according to Eq. (\ref{commut}), coincides with algebra of Pauli matrices $ \left[ \hat{\sigma}_{\alpha},\hat{\sigma}_{\beta} \right] = 2i \hat{\sigma}_{\gamma} $ ($\alpha,\beta,\gamma = 1,2,3 $) at   $ \hat{\mathcal{I}}_{\alpha} \rightarrow \hat{\sigma}_{\alpha} $. 
This explains the meaning of notations introduced in Eq. (\ref{G_1,2,3}). 

Thus, we can introduce new representation for matrices  $ \hat{\mathcal{K}} $, $ \hat{\mathcal{A}} $ and $ \hat{\mathcal{I}} $:  
\begin{equation}
\label{repBAK} 
\hat{\mathfrak{K}} = \sqrt{\hat{G}_{1}\left( \hat{\mathcal{J}}^{2} \right)} \otimes \hat{\sigma}_{1} , \quad \hat{\mathfrak{A}} = \sqrt{\hat{G}_{2}\left( \hat{\mathcal{H}}, \hat{\mathcal{J}}^{2} \right)} \otimes \hat{\sigma}_{2} , \quad \hat{\mathfrak{I}_{BEL}} = \sqrt{\hat{G}_{3}\left(\mathcal{H}, \hat{\mathcal{J}}^{2} \right)} \otimes \hat{\sigma}_{3} .
\end{equation}
Operators $ \hat{\mathcal{H}} $, $ \hat{\mathcal{J}}^{2} $ and $ \hat{\mathcal{J}}_{z} $ in this representation have the form
$$ \hat{\mathfrak{H}} = \hat{\mathcal{H}} \otimes \hat{\sigma}_{0} , \quad \hat{\mathfrak{J}}^{2} = \hat{\mathcal{J}}^{2} \otimes \hat{\sigma}_{0} , \quad \hat{\mathfrak{J}}_{z} = \hat{\mathcal{J}}_{z} \otimes \hat{\sigma}_{0} , $$ 
where $ \hat{\sigma}_{0} $ is a unit $ 2 \times 2 $ matrix. 

The field of the bispinors in this new 'spinor' representation is defined as 
\begin{equation}
\label{statevec} 
 \vert \varepsilon ,j, m_{j},\sigma \rangle_{inv} = \left( \begin{array}{c}
 \vert \varepsilon ,j, m_{j} \rangle_{u} \\
 \vert \varepsilon ,j, m_{j} \rangle_{d}
\end{array} \right)_{inv} , 
\end{equation}
where the upper, $ u, $ and lower, $ d, $ spinors satisfy the first three equalities in the system of equations (\ref{state_3}) and are independent of the invariant. The operators $ \hat{G}_{\alpha } $ in the space of these bispinors are reduced to their eigenvalues (\ref{eigenG_123}), namely, $ \hat{G}_{\alpha} \rightarrow g_{\alpha} $, or $ g_{1} = \kappa_{j}^{2}  $, $ g_{2} = a_{\varepsilon,j}^{2} $, $ g_{3} =\kappa_{j}^{2} a_{\varepsilon,j}^{2} $. 

There is a one-to-one correspondence between the elements of equalities (\ref{state_3}) 
  and their presentation in the form defined by Eqs. (\ref{repBAK})-(\ref{statevec}), so that  Eqs. (\ref{state_3}), (\ref{commut}) and (\ref{C^2}) are preserved. Each component of the spinor 
  (\ref{statevec}) as of the geometrical object determines the direction in three-dimensional space depending on  the orientation of the coordinate system. If the latter is fixed, spinors $ \chi_{\uparrow} = \left( 1 \: 0 \right)^{T} $ and $ \chi_{\downarrow} = \left( 0 \: 1 \right)^{T} $ indicate directions along and opposite to one of the orthogonal axes, so that  $ \chi_{\uparrow} $ and $ \chi_{\downarrow} $ are eigen spinors of the diagonal Pauli matrix $ \hat{\sigma}_{3} $. That is why the index $ inv $ in the representation (\ref{statevec}) shows the orientation of the "coordinate system" which is defined depending on which invariant from the set   (\ref{repBAK}) the diagonal Pauli matrix  $ \hat{\sigma}_{3} $ corresponds to. 

As the initial orientation it is convenient to choose the one, for which in the 'spinor' representation (see (\ref{calC})) 
\[
\hat{\mathfrak{C}} = 
\hat{\mathfrak{K}} + \hat{\mathfrak{A}} + \hat{\mathfrak{I}} = \sum_{\alpha =1}^3 \sqrt{\hat{G}_{\alpha}} \otimes \hat{\sigma}_{\alpha} 
\]
matrix  $ \hat{\sigma}_{1} $ is diagonal. In the spacial notations  $ \hat{\sigma}_{1} =  \hat{\sigma}_{z} $, $ \hat{\sigma}_{2} =  \hat{\sigma}_{y} $ and $ \hat{\sigma}_{3} =  \hat{\sigma}_{x} $ one can obtain
\begin{equation}
\label{reprK} 
\hat{\mathfrak{C}} = \left( \begin{array}{cc}
0 & \sqrt{\hat{G}_{2}} \\
\sqrt{\hat{G}_{2}} & 0 
\end{array} \right) + \left( \begin{array}{cc}
0 & -i \sqrt{\hat{G}_{3}} \\
i \sqrt{\hat{G}_{3}} & 0 
\end{array} \right) + \left( \begin{array}{cc}
\sqrt{\hat{G}_{1}} & 0 \\
0 & -\sqrt{\hat{G}_{1}} 
\end{array} \right) .
\end{equation}

As it has been mentioned above, the system of eigen vectors  $ \vert \varepsilon ,j, m_{j},\sigma \rangle_{inv} $ depends on the choice of the invariant in the fourth equation in system (\ref{state_3}), therefore, the states with the definite value of the invariant  $ \hat{K} $ can be found from the equation  $ \hat{\mathfrak{K}} \vert \varepsilon ,j, m_{j} \rangle  = \epsilon_{\mathcal{K}} \vert \varepsilon ,j, m_{j} \rangle $, whose explicit form 
\[
\left( \begin{array}{cc}
\sqrt{\hat{G}_{1}} & 0 \\
0 & -\sqrt{\hat{G}_{1}} 
\end{array} \right) \left( \begin{array}{c}
 \vert \varepsilon ,j, m_{j} \rangle_{u} \\
 \vert \varepsilon ,j, m_{j} \rangle_{d}
\end{array} \right)_{(\mathcal{K})} = \epsilon_{\mathcal{K}}  \left( \begin{array}{c}
 \vert \varepsilon ,j, m_{j} \rangle_{u} \\
 \vert \varepsilon ,j, m_{j} \rangle_{d}
\end{array} \right)_{(\mathcal{K})}
\]
allows one to find easily two states $ \vert \varepsilon ,j, m_{j},\sigma \rangle_{\mathcal{K}} $ which correspond to eigenvalues  $ \epsilon_{\mathcal{K}} = \sigma \kappa_{j} $, where $ \sigma = \pm $, namely,
\begin{equation}
\label{Kstates} 
\vert \varepsilon ,j, m_{j},+ \rangle_{\mathcal{K}} = \left( \begin{array}{c}
\vert \varepsilon ,j, m_{j} \rangle_{u} \\
0
\end{array} \right)_{(\mathcal{K})} , \quad \vert \varepsilon ,j, m_{j},- \rangle_{\mathcal{K}} = \left( \begin{array}{c}
0 \\
\vert \varepsilon ,j, m_{j} \rangle_{d}
\end{array} \right)_{(\mathcal{K})} .
\end{equation}

The states with definite value of the invariant  $ \hat{A} $ are defined, respectively, by the equation $ \hat{\mathfrak{A}} \vert \varepsilon ,j, m_{j} \rangle  = \epsilon_{\mathcal{A}} \vert \varepsilon ,j, m_{j} \rangle $. Although matrix  $ \hat{\mathfrak{A}} $ is non-diagonal, its eigenvalues can be found, since it can be reduced to the diagonal form by rotation   
\begin{equation}
\label{R_j} 
 \hat{\mathcal{R}}_{\alpha} \left( \phi \right) = e^{i(\phi /2) \hat{\sigma}_{\alpha}} = \cos \frac{\phi}{2} + i \sin \frac{\phi}{2} \hat{\sigma}_{\alpha} , \quad \alpha = x,y,z   
\end{equation}
of the "coordinate system" in the three-dimensional abstract space around its orthogonal axes. In particular,  operation   $ \hat{\mathcal{R}}_{x} \left( -\pi /2 \right) $ transforms vector (\ref{reprK}) to the form
\begin{equation}
\label{R_x1} 
\hat{\mathcal{R}}_{x} \left( -\pi /2 \right) \hat{\mathfrak{C}} \hat{\mathcal{R}}_{x}^{\dagger} \left( -\pi /2 \right) = \hat{\tilde{\mathfrak{C}}} = \sqrt{\hat{G}_{3}} \otimes \hat{\sigma}_{x} - \sqrt{\hat{G}_{1}} \otimes \hat{\sigma}_{y} + \sqrt{\hat{G}_{2}} \otimes \hat{\sigma}_{z} .
\end{equation}

In the "coordinate system" rotated in such a way, the invariant $ \hat{\tilde{\mathfrak{A}}} $ becomes diagonal, and the equation 
 $ \hat{\tilde{\mathfrak{A}}} \vert \tilde{\psi} \rangle_{\mathcal{A}} = \epsilon_{\mathcal{A}} \vert \tilde{\psi} \rangle_{\mathcal{A}} $ determines states  $ \vert \varepsilon ,j, m_{j},\sigma \rangle_{\mathcal{A}} $ with eigenvalues $ \epsilon_{\mathcal{A}} = \sigma a_{\varepsilon,j} $ with the same values of $ \sigma = \pm $. Here  $ a_{\varepsilon,j} $ is defined by Eq. (\ref{a_E,j}) (see also (\ref{eigenG_123})). In the result, we come to vectors, similar to the ones in  Eq. (\ref{Kstates}): 
\[
\vert \varepsilon ,j, m_{j},+ \rangle_{\mathcal{A}} = \left( \begin{array}{c}
\vert \varepsilon ,j, m_{j} \rangle_{u} \\
0
\end{array} \right)_{(\mathcal{A})} , \quad \vert \varepsilon ,j, m_{j},- \rangle_{\mathcal{A}} = \left( \begin{array}{c}
0 \\
\vert \varepsilon ,j, m_{j} \rangle_{d}
\end{array} \right)_{(\mathcal{A})} .
\]

The states in rotated and initial spaces are also related by rotation (\ref{R_j}) on angle $ \phi = -\pi/2 $:
\[ 
\vert \tilde{\psi} \rangle_{\mathcal{A}} = \hat{\mathcal{R}}_{x} \left( -\pi /2 \right) \vert \psi \rangle_{\mathcal{K}} , \quad \vert \tilde{\psi} \rangle_{\mathcal{K}} = \hat{\mathcal{R}}^{\dagger}_{x} \left( -\pi /2 \right) \vert \psi \rangle_{\mathcal{A}} . 
\]
Taking into account the latter relation,  solution of the DE corresponding to Johnson-Lippmann invariant, in the initial coordinate system has the form  
\begin{equation}
\label{solA-K} 
\begin{array}{c}
\vert \varepsilon ,j, m_{j},+ \rangle_{\mathcal{A}} = e^{i\pi /4} \frac{1}{\sqrt{2}} \vert \varepsilon ,j, m_{j},+ \rangle_{\mathcal{K}} -  e^{-i\pi /4} \frac{1}{\sqrt{2}} \vert \varepsilon ,j, m_{j},- \rangle_{\mathcal{K}}  , \\ 
\vert \varepsilon ,j, m_{j},- \rangle_{\mathcal{A}} = e^{i\pi /4} \frac{1}{\sqrt{2}} \vert \varepsilon ,j, m_{j},+ \rangle_{\mathcal{K}} + e^{-i\pi /4} \frac{1}{\sqrt{2}} \vert \varepsilon ,j, m_{j},- \rangle_{\mathcal{K}}  .
\end{array} 
\end{equation}

Finally, at the rotation  $ \hat{\mathcal{R}}_{y} \left( \pi /2 \right) $, vector (\ref{calC}) is transformed to the combination  (cf. (\ref{R_x1})) 
\[
\hat{\tilde{\mathfrak{C}}} = \hat{\mathcal{R}}_{y} \left( \pi /2 \right) \hat{\mathfrak{C}} \hat{\mathcal{R}}^{\dagger}_{y} \left( \pi /2 \right) = \sqrt{\hat{G}_{1}} \otimes \hat{\sigma}_{x} + \sqrt{\hat{G}_{2}} \otimes \hat{\sigma}_{y} + \sqrt{\hat{G}_{3}} \otimes \hat{\sigma}_{z} ,
\]
when  the component  $ \hat{\mathfrak{I}}_{BEL} $ is diagonal. It has two eigenvalues of the opposite sign  $ \epsilon_{\mathcal{I}} = \sigma b_{\varepsilon,j} $, where again  $ \sigma = \pm $. Therefore,  index $ \sigma = \pm $ distinguishes the eigenstates $ \vert \tilde{\psi} \rangle_{\mathcal{I}} = \vert \varepsilon ,j, m_{j},\sigma \rangle_{\mathcal{I}} $ with $\sigma =+$ and $\sigma =-$. In this case using the transformation to the initial coordinate system, the solution of the DE with the definite value of the invariant  $ \hat{\mathcal{I}} $ can be represented in the form (cf. (\ref{Kstates}) and (\ref{solA-K})) 
\begin{equation}
\label{solB-K} 
\begin{array}{c}
\vert \varepsilon ,j, m_{j},+ \rangle_{\mathcal{I}} = \frac{1}{\sqrt{2}} \vert \varepsilon ,j, m_{j},+ \rangle_{\mathcal{K}} + \frac{1}{\sqrt{2}} \vert \varepsilon ,j, m_{j},- \rangle_{\mathcal{K}}  , \\ 
\vert \varepsilon ,j, m_{j},- \rangle_{\mathcal{I}} = - \frac{1}{\sqrt{2}} \vert \varepsilon ,j, m_{j},+ \rangle_{\mathcal{K}} + \frac{1}{\sqrt{2}} \vert \varepsilon ,j, m_{j},- \rangle_{\mathcal{K}} .
\end{array} 
\end{equation}

In the general case of an arbitrary rotation  $ \hat{\mathcal{R}} \left( \psi ,\theta ,\phi \right) = \hat{\mathcal{R}}_{z} \left( \psi \right) \hat{\mathcal{R}}_{x} \left( \theta \right) \hat{\mathcal{R}}_{z} \left( \phi \right) $ which is defined by the Euler angles $ \phi $, $ \theta $ and $ \psi $, vector  (\ref{reprK}) is transformed in any new "coordinate system" according to the rule  $ \hat{\tilde{\mathfrak{C}}} = \hat{\mathcal{R}} \left( \psi ,\theta ,\phi \right) \hat{\mathfrak{C}} \hat{\mathcal{R}}^{\dagger} \left( \psi ,\theta ,\phi \right) $. In such a case the expression for the diagonal component of vector $ \hat{\tilde{\mathfrak{C}}} $, which is the integral of motion, has relatively simple form 
\[
\hat{\mathfrak{I}}_{gen} = \left( \sqrt{\hat{G}_{1}} \cos \theta - \sqrt{\hat{G}_{2}} \cos \phi \sin \theta + \sqrt{\hat{G}_{3}} \sin \phi \sin \theta \right) \otimes \hat{\sigma}_{z} .
\] 
In the space of states $ \vert \varepsilon ,j, m_{j} \rangle $ this component, as can be easily checked, has two eigenvalues,  $ \epsilon_{gen} = \pm \left( \kappa_{j} \cos \theta - a_{\varepsilon,j} \cos \phi \sin \theta + \kappa_{j} a_{\varepsilon,j} \sin \phi \sin \theta \right) $. Together with the energy and total momentum, it takes certain value, and, thus, is the generalized invariant 
\[
\hat{\mathcal{I}}_{gen} = C_{\mathcal{K}} \hat{\mathcal{K}} + C_{\mathcal{A}} \hat{\mathcal{A}} + C_{\mathcal{I}} \hat{\mathcal{I}} , \quad C_{\mathcal{K}} = \cos \theta , \; C_{\mathcal{A}} = -\cos \phi \sin \theta , \; C_{\mathcal{I}} = \sin \phi \sin \theta ,
\]
which has been introduced in \cite{AoP22}. 

 The states(\ref{eignvecs_n})  with  $ n_{r} = 0 $ and energy   (\ref{E_n}) are very special. Corresponding to them eigenvalues of the operators $ \hat{\mathcal{A}} $ and $ \hat{\mathcal{I}} $ are equal  zero, according to Eq. (\ref{eqE_n}). 
 %Since, in similar particular cases eigenvalues of these operators are zero, 
 Therefore, the sign $ \sigma = \pm $ for these states has no meaning, the two solutions are reduced to a single one 
$$ \vert n,0,m_{j},+ \rangle_{\mathcal{A}} = \vert n,0,m_{j},- \rangle_{\mathcal{A}} = \vert n,0,m_{j} \rangle_{\mathcal{A}} , $$  
where the states  $ \vert n,n_{r},m_{j},\sigma \rangle_{\mathcal{A}} $ are defined in (\ref{solA-K}). Adding and subtracting these two expressions at  $ n_{r} = 0 $, one comes to the relations 
\[
2 \vert n,0,m_{j} \rangle_{\mathcal{A}} = e^{i\pi /4} \frac{2}{\sqrt{2}} \vert n,0,m_{j},+ \rangle_{\mathcal{K}} , \quad 0 = e^{-i\pi /4} \frac{2}{\sqrt{2}} \vert n,0,m_{j},- \rangle_{\mathcal{K}} ,
\]
which demonstrate that $ \vert n,0,m_{j},- \rangle_{\mathcal{K}} = 0 $.

Therefore, the states with quantum numbers $ \lbrace n,0,m_{j} \rbrace $ or $ \lbrace n,j_{max},m_{j} \rbrace $ are described by the bispinor
\begin{equation}
\label{state_r=0} 
\vert n,j_{max},m_{j} \rangle = \left( \begin{array}{c}
\vert n,j_{max},m_{j} \rangle_{u} \\
0
\end{array} \right)
\end{equation}
independent on the choice of the spinor invariant. In the consequence, in the system of eigenstates with the  definite value of the Dirac invariant at   $ n_{r} = 0 $, the vector with $ \sigma = + $ is determined by the expression (\ref{state_r=0}), while the state with $ \sigma = - $ turns out to be equal to zero, in spite of the fact that its eigenvalue  $ \hat{\mathcal{K}} $ is non-zero ($ \kappa_{j} = n $ at $ n_{r} = 0 $). The same conclusion is preserved when the DE is solved directly in  spherical coordinates  \cite{AoP22}. 

Notice that among the operations  $ \hat{\mathcal{R}} \left( \psi ,\theta ,\psi \right) $ the rotation on the angles $ \phi = \phi_{0} $ and $ \theta = \theta_{0} $, defined by equalities 
\begin{equation}
\label{phi,teta_0} 
\tan \phi_{0} = - \frac{\sqrt{\hat{G}_{3}}}{\sqrt{\hat{G}_{2}}} , \quad \tan \theta_{0} = - \frac{\sqrt{\hat{G}_{3} + \hat{G}_{2}}}{\sqrt{\hat{G}_{1}}} ,
\end{equation}
plays a special role. In this rotated "coordinate system" matrix   $ \hat{\mathfrak{\tilde{C}}} $ reduces to the expression 
\begin{equation}
 \label{C_eig} 
\hat{\mathfrak{\tilde{C}}} = \hat{\mathcal{R}}_{z} \left( \psi \right)  \sqrt{\hat{G}_{1} + \hat{G}_{2} + \hat{G}_{3}} \otimes \hat{\sigma}_{z}  \hat{\mathcal{R}}^{\dagger}_{z} \left( \psi \right) = \hat{\mathcal{C}}_{\rm{eig}} = \sqrt{\hat{G}_{1} + \hat{G}_{2} + \hat{G}_{3}} \otimes \hat{\sigma}_{z} . 
 \end{equation} 
It follows that the rotation  $ \hat{\mathcal{R}}_{x} \left( \theta_{0} \right) \hat{\mathcal{R}}_{z} \left( \phi_{0} \right) $ transforms the matrix $ \hat{\mathfrak{C}} $ (\ref{reprK}) to the diagonal form, when the chosen "axis" ($ "z" $) coincides with the eigen axis of the vector. In such a case matrix   $ \hat{\mathcal{C}}_{eig} $ is invariant with respect to the rotation $ \hat{\mathcal{R}}_{z} \left( \psi \right) $ by the angle $ \psi $ around the axis, which coincides with the direction of the vector associated with it. In the states $ \hat{\mathfrak{\tilde{C}}} $ as in a particular case of the generalized invariant, operator (\ref{C_eig}) has two eigenvalues with the opposite signs 
\[
c_{ \varepsilon ,j}^{2} = \pm 
\sqrt{\epsilon_{\mathcal{K}}^2+ \epsilon_{\mathcal{A}} ^2+ \epsilon_{\mathcal{I}}^2},
\]
which distinguish the corresponding states  $ \vert \varepsilon ,j, m_{j},\sigma \rangle_{\mathcal{C}} $.

\section{Conclusion}

In the present paper problem of the Dirac electron in the Coulomb field (relativistic Kepler problem) has been studied. It has been shown that the properties and peculiarities of the electron stationary states and corresponding eigenvectors from the Hilbert space can be obtained within the algebraic approach based on the algebra of the three spinor invariants   $ \hat{K} $, $ \hat{A} $, $ \hat{I} $, which relativistically incorporate spin degree of freedom (see Eqs. (\ref{Ksc}), (\ref{invJ-L}), (\ref{invBEL}), respectively). It has been shown that the stationary states $ \vert \varepsilon ,j, m_{j},\sigma \rangle_{inv} $, according to Eq.(\ref{eignvecs_n}), are naturally characterized by the quantum numbers   $ \lbrace n,j,m_{j},\sigma \rbrace $, where $ n $ is the principal quantum number. When it is fixed, the orbital quantum number  $ j $ can attain $ n $ values which are bound by the maximal value $ j_{max} $, for which   $ \kappa_{j_{max}} = n $. The states with  $ j < j_{max} $ correspond to the orbital number $ j = j_{max} - n_{r} $, where the number $ n_{r} = 0,1,\ldots \, , n-1 $ takes $ n $ values and has the meaning of the radial quantum number. The symbol $ \sigma $ in the set of quantum numbers takes two values,  $ \sigma = \pm $, and determines the eigenvalue sign of the spinor invariant operator. It turns out that in the states with zero  radial number, $ n_{r} = 0 $, in which the eigenvalues of two invariants  $ \hat{A} $ and $ \hat{I_{BEL}} $ are zero, the energy value is given by Eq. (\ref{E_n}), the discrete index $ \sigma $ has one value, only, and the corresponding eigenvector is given by Eq. (\ref{state_r=0}) independent on the choice of a particular spinor invariant . 

Solutions of the DE in the coordinate space are given by the bispinors $ \vert \varepsilon ,j, m_{j}, \kappa  \rangle_{inv} $ which depend on concrete spinor invariant, as it has been shown above. The states $ \vert \varepsilon ,j, m_{j},\sigma \rangle_{\mathcal{K}} $ with certain values of the Dirac invariant are given in the coordinate space by two Darwin solutions \cite{Darwin,Gordon}, while the states $ \vert \varepsilon ,j, m_{j},\sigma \rangle_{\mathcal{A}} $ and $ \vert \varepsilon ,j, m_{j},\sigma \rangle_{\mathcal{B}} $ correspond to the solutions of the DE with certain values of the invariants   $ \hat{\mathcal{A}} $ and $ \hat{\mathcal{B}} $, respectively, which were found in  \cite{AoP22}. The relation between different solutions  (\ref{solA-K}) and (\ref{solB-K}), calculated  here algebraically approach, coincides with the relation obtained by direct solving of the DE in the coordinate space in \cite{AoP22}.  

Another important conclusion, in our opinion, is related with the invariance of  results with respect to the rotation around  axis which coincides with the direction of  vector (\ref{C_eig}) that is associated with  matrix $ \hat{\mathcal{C}} $. This invariance indicates dynamical (hidden) symmetry $ SU(2) $.

In the initial "coordinate system" (\ref{reprK}) the rotation operator $\hat{R} \left( \psi \right)$ in Eq. (\ref{C_eig}) takes the form
\[
\hat{R} \left( \psi \right) = \hat{\mathcal{R}}^{\dagger}_{z}\left( \phi_{0} \right) \hat{\mathcal{R}}^{\dagger}_{x} \left( \theta_{0} \right) \hat{\mathcal{R}}_{z} \left( \psi \right) \hat{\mathcal{R}}_{z}\left( \phi_{0} \right) \hat{\mathcal{R}}_{x} \left( \theta_{0} \right) = \exp \left( i\frac{\psi}{2} \hat{S} \right) ,
\]
where
\[
\hat{S} = \hat{\mathcal{R}}^{\dagger}_{z}\left( \phi_{0} \right) \hat{\mathcal{R}}^{\dagger}_{x} \left( \theta_{0} \right) \hat{\sigma}_{z} \hat{\mathcal{R}}_{z}\left( \phi_{0} \right) \hat{\mathcal{R}}_{x} \left( \theta_{0} \right) = \frac{\hat{\mathcal{C}}}{\sqrt{\hat{G}_{1} + \hat{G}_{2} + \hat{G}_{3}}} , \quad \hat{S}^{2} = \hat{I} ,
\]
and angles $ \phi_{0} $ and $ \theta_{0} $ are defined by Eq. (\ref{phi,teta_0}). Operator  $ \hat{R} \left( \psi \right) $ commutes with the Hamiltonian, and since the equality $ \hat{S}^{2} = \hat{\mathrm{I}} $ is valid, it has the form 
$$ \hat{R} \left( \psi \right) = \cos \frac{\psi}{2} \hat{\mathrm{I}} + i \sin \frac{\psi}{2} \hat{S} = 
\left( \begin{array}{cc}
(c_1) \left( \psi \right) & -(c_2)^{\ast} \left( \psi \right) \\
(c_2) \left( \psi \right) & (c_1)^{\ast} \left( \psi \right) , 
\end{array} \right) . $$
Transformation coefficients $ c_1  $ and $ c_2 $ in the bispinor field are complex, and the action of this operator on eigenstates
(\ref{statevec}) with the components $ \vert \Psi_{n,j,m_{j} ,+} \rangle $ and $ \vert \Psi_{n,j,m_{j} ,-} \rangle $ that are the solutions of the DE, transforms them to the new solution 
\[
\begin{array}{c}
\vert \tilde{\Psi}_{n,j,m_{j} ,+} \rangle = (c_1) (\left( \psi \right) \vert \Psi_{n,j,m_{j} ,+} \rangle - (c_2)^{\ast} \left( \psi \right) \vert \Psi_{n,j,m_{j} ,-} \rangle , \\ 
\vert \tilde{\Psi}_{n,j,m_{j} ,-} \rangle = (c_2) \left( \psi \right) \vert \Psi_{n,j,m_{j} ,+} \rangle + (c_1)^{\ast} \left( \psi \right) \vert \Psi_{n,j,m_{j} ,-} \rangle ,
\end{array}
\]
which has been shown actually in  \cite{AoP22}. Namely the Hamiltonian invariance with respect to this transformation indicates its $ SU(2) $ symmetry. The complete symmetry of the DE is described by the $ SO(3) \otimes SU(2) $ group. The generator of  $ SO(3) $ group is given by the total momentum operator $ \hat{\mathbf{J}} $, and the generator of  $ SU(2) $ group is given by the rotation of bispinors in the spinor space, determined by the invariants  $ \hat{K} $, $ \hat{A} $ and $ \hat{I_BEL} $. 

\vskip5mm
We declare the absence of the conflict of interest. 

\vskip5mm 
{\bf Acknowledgement.} 
\textit{This work was supported by the fundamental scientific program 0122U000887 of the Department of Physics and Astronomy of the National Academy of Sciences of Ukraine. We acknowledge also Grant 0122U002313 of the National Academy of Sciences of Ukraine }

\end{document}